\documentclass[aps,pra,twocolumn,superscriptaddress]{revtex4-2}
\pdfoutput=1

\usepackage[utf8]{inputenc}
\usepackage[english]{babel}
\usepackage{amsfonts, amsmath, amssymb, amsthm}
\usepackage{graphicx}
\usepackage{xcolor}
\usepackage{bm}
\usepackage{braket}
\usepackage[noend]{algpseudocode}
\usepackage{algorithm}
\usepackage{algorithmicx}
\usepackage[normalem]{ulem}
\usepackage[colorlinks=true,breaklinks=true,linkcolor=black,citecolor=green,urlcolor=magenta]{hyperref}
\usepackage[capitalise]{cleveref}

\algnewcommand\algorithmicinput{\textbf{Input:}}
\algnewcommand\Input{\item[\algorithmicinput]}

\algnewcommand\algorithmicoutput{\textbf{Output:}}
\algnewcommand\Output{\item[\algorithmicoutput]}

\usepackage{bm}
\usepackage{braket}

\newcommand\device[1]{$\mathsf{ibm}\_\mathsf{#1}$}
\newcommand\crt[1]{\hat{a}^\dagger_{#1}}
\newcommand\dst[1]{\hat{a}^{\phantom{\dagger}}_{#1}}
\newcommand\bts[1]{{\bf{#1}}}
\newcommand\scientific[2]{#1 \cdot 10^{#2}}
\newcommand\excitation[2]{\hat{E}^{#1}_{#2}}
\newcommand\captiontitle[1]{{\bf{#1}}.\,}
\usepackage{xcolor}
\newcommand{\nocontentsline}[3]{}
\newcommand{\tocless}[2]{\bgroup\let\addcontentsline=\nocontentsline#1{#2}\egroup}

\begin{document}

\title{Towards quantum-centric simulations of extended molecules: sample-based quantum diagonalization enhanced with density matrix embedding theory}

\author{Akhil Shajan}
\affiliation{Center for Computational Life Sciences, Lerner Research Institute, The Cleveland Clinic, Cleveland, Ohio 44106, United States}
\affiliation{Department of Chemistry, Michigan State University, East Lansing, Michigan 48824, United States}

\author{Danil Kaliakin}
\affiliation{Center for Computational Life Sciences, Lerner Research Institute, The Cleveland Clinic, Cleveland, Ohio 44106, United States}

\author{Abhishek Mitra}
\affiliation{Center for Computational Life Sciences, Lerner Research Institute, The Cleveland Clinic, Cleveland, Ohio 44106, United States}

\author{Javier Robledo Moreno}
\affiliation{IBM Quantum, IBM T.J. Watson Research Center, Yorktown Heights, NY 10598, United States}

\author{Zhen Li}
\affiliation{Center for Computational Life Sciences, Lerner Research Institute, The Cleveland Clinic, Cleveland, Ohio 44106, United States}

\author{Mario Motta}
\affiliation{IBM Quantum, IBM T.J. Watson Research Center, Yorktown Heights, NY 10598, United States}

\author{Caleb Johnson}
\affiliation{IBM Quantum, IBM T.J. Watson Research Center, Yorktown Heights, NY 10598, United States}

\author{Abdullah Ash Saki}
\affiliation{IBM Quantum, IBM T.J. Watson Research Center, Yorktown Heights, NY 10598, United States}

\author{Susanta Das}
\affiliation{Center for Computational Life Sciences, Lerner Research Institute, The Cleveland Clinic, Cleveland, Ohio 44106, United States}

\author{Iskandar Sitdikov}
\affiliation{IBM Quantum, IBM T.J. Watson Research Center, Yorktown Heights, NY 10598, United States}

\author{Antonio Mezzacapo}
\affiliation{IBM Quantum, IBM T.J. Watson Research Center, Yorktown Heights, NY 10598, United States}

\author{Kenneth M. Merz Jr.}
\email{kmerz1@gmail.com}
\affiliation{Center for Computational Life Sciences, Lerner Research Institute, The Cleveland Clinic, Cleveland, Ohio 44106, United States}
\affiliation{Department of Chemistry, Michigan State University, East Lansing, Michigan 48824, United States}


\begin{abstract}
Computing ground-state properties of molecules is a promising application for quantum computers operating in concert with classical high-performance computing resources.
Quantum embedding methods are a family of algorithms particularly suited to these computational platforms: they combine high-level calculations on active regions of a molecule with low-level calculations on the surrounding environment, thereby avoiding expensive high-level full-molecule calculations and allowing to distribute computational cost across multiple and heterogeneous computing units.
Here, we present the first density matrix embedding theory (DMET) simulations performed in combination with the sample-based quantum diagonalization (SQD) method.
We employ the DMET-SQD formalism to compute the ground-state energy of a ring of 18 hydrogen atoms, and the relative energies of the chair, half-chair, twist-boat, and boat conformers of cyclohexane.
The full-molecule 41- and 89-qubit simulations are decomposed into 27- and 32-qubit active-region simulations, that we carry out on the \device{cleveland} device, obtaining results in agreement with reference classical methods.
Our DMET-SQD calculations mark a tangible progress in the size of active regions that can be accurately tackled by near-term quantum computers, and are an early demonstration of the potential for quantum-centric simulations to accurately treat the electronic structure of large molecules, with the ultimate goal of tackling systems such as peptides and proteins.
\end{abstract}

\maketitle

\tocless\section{Introduction}

The accurate treatment of interacting many-electron systems by first-principle computational methods is a grand challenge of contemporary science.
Progress in addressing this challenge will affect diverse fields, including pharmaceutical research and in particular computer-aided drug design, where it is important to characterize the energy landscape of the conformations of large molecule and their complexes at room temperature~\cite{bardwell2011towards,yang2014ab}.
Existing computational resources, however, are insufficient to carry out first-principle electronic structure simulations on many large molecules, e.g. proteins, by methods beyond the mean-field approximation.
For example, the simulation of insulin~\cite{timofeev2010x}  using minimal and correlation-consistent cc-pVxZ (x=D,T,Q) basis sets requires 2418 and 7561, 17420, 33559 molecular orbitals (MOs) respectively, at the boundaries of conventional electronic structure methods~\cite{sun2020recent, valiev2010nwchem}.

Some approximations, therefore, are needed to realize quantum mechanical calculations on such systems. An example is offered by the large family of fragment-based electronic structure methods~\cite{kitaura1999fragment, raghavachari2015accurate, erlanson2019fragment}.
These algorithms are based on the decomposition of an intractably large system into tractable subsystems, and typically employ the quantum embedding scheme, that combines the treatment of subsystems by a higher-level method (i.e. more accurate and expensive) with a lower-level computation on the entire system~\cite{sun2016quantum}. They thus offer an important path to leverage advances in first-principle electronic structure methods in the modeling of otherwise intractably large molecules.

Quantum embedding methods are also natural and compelling targets for quantum-centric supercomputing (QCSC) architectures~\cite{alexeev2024quantum}, where a quantum computer operates in concert with classical high-performance computing (HPC) resources.
In this framework, high-level quantum computations can be carried out on subsystems, with HPC resources providing feedback between the subsystems and their environment. 
Various authors have recognized this possibility, and have proposed to integrate quantum computing subroutines in the workflow of quantum embedding methods~\cite{bauer2016hybrid, kreula2016few, bravyi2017complexity} where the feedback between subsystems and their environment is based on the electronic density, the Green's function, or the one-particle reduced density matrix (depending on the specific embedding theory)~\cite{sun2016quantum}.
Before the availability of large-scale fault-tolerant quantum computers, a critical aspect of the application of QCSC to quantum embedding is the optimization of the resources of near-term quantum computers, limited in both qubit number and quantum circuit size.
Motivated by this consideration, several authors have proposed to use the variational quantum eigensolver (VQE) method~\cite{peruzzo2014variational, motta2022emerging, bauer2020quantum} in combination with the fragment molecular orbital (FMO)~\cite{lim2024fragment}, divide and conquer (DC)~\cite{yoshikawa2022quantum}, many-body expansion (MBE)~\cite{ma2023multiscale}, density matrix embedding theory (DMET)~\cite{rubin2016hybrid, kawashima2021optimizing, iijima2023towards, li2022toward, kirsopp2022quantum, cao2023ab, ma2023multiscale}, and other wavefunction-based embedding~\cite{vorwerk2022quantum, rossmannek2023quantum, gujarati2023quantum} methods.

However, there are some drawbacks that are inevitable upon introducing fragmentation approximations, that are exacerbated by the constrained size of subsystems accurately tractable by near-term quantum algorithms and computers: for example, the dispersion and induction of many-body interactions and the cooperativity of hydrogen bonds are considerably impacted. Therefore, extending the reach of quantum computers in terms of tractable subsystem sizes and accuracy of results stands to mitigate the impact of the fragmentation approximation, allowing more meaningful QCSC applications in the framework of quantum embedding.

The recently introduced sample-based quantum diagonalization (SQD) method~\cite{robledo2024chemistry} has allowed electronic structure simulations for active sites of metallo-proteins using up to 77 qubits (corresponding to 36 MOs), a substantial advancement in application of quantum computers to quantum chemistry~\cite{guo2024experimental, zhao2023orbital, google2020hartree, huang2023leveraging}, and for molecular dimers bound by non-covalent interactions~\cite{kaliakin2024accurate}, as well as the large active space simulations of methylene~\cite{Ieva2024}. In view of these results, it is timely and compelling to investigate the use of SQD as a subsystem solver in the framework of the well-established DMET quantum embedding method. Herein, we present the first DMET-SQD simulations of molecular systems. We use DMET-SQD to compute the ground-state potential energy curve of a ring of 18 hydrogen atoms, a well-established benchmark test~\cite{hachmann2006multireference, stella2011strong, motta2017towards, motta2020ground}. We then compute the relative energies of the chair, half-chair, twist-boat, and boat conformers of cyclohexane, the structure and dynamics of which are important prototypes of a wide range of organic compounds~\cite{barton1956principles, strauss1970conformational, dixon1990ab, eliel1972conformational}. Our calculations use 27 and 32 qubits for the hydrogen ring and cyclohexane respectively, a substantial reduction from the 41 and 89 qubits required for unfragmented calculations, and are carried out on the \device{cleveland} quantum computer from IBM's Eagle processor family.
To assess the accuracy of our results, we compare them against DMET calculations using exact diagonalization (FCI) as a subsystem solver -- to estimate the accuracy of SQD as a subsystem solver -- and unfragmented calculations with heat-bath configuration interaction (HCI) and coupled cluster singles and doubles (CCSD) -- to quantify the accuracy of the fragmentation approximation.

\vspace{2em}
\tocless\section{Methods}

\subsection*{Density matrix embedding theory}

DMET~\cite{knizia2012density, knizia2013density, wouters2016practical, wouters2017five, pham2018can} is a wave-function-based quantum embedding method, that allows high-level treatments for multiple subsystems using accurate and expensive beyond mean-field methods in conjunction with a mean-field calculation on the entire system. Formally, DMET divides the $D$-dimensional Hilbert space of a quantum system into a fragment, i.e. a subspace of dimension $d_F \ll D$, and an environment, i.e. a subspace of dimension $D_E = D-d_F \gg d_F$. The exact and unknown ground-state of the system's Hamiltonian $\hat{H}$, $| \Psi \rangle = \sum_{f=1}^{d_F} \sum_{e=1}^{D_E} \Psi_{fe} | f \rangle \otimes | e \rangle$, is written through a Schmidt decomposition of the matrix $\Psi_{fe} = \sum_{\mu=1}^{d_F} U_{f\mu} \lambda_\mu V_{\mu e}$ as $| \Psi \rangle = \sum_{\mu=1}^{d_F} \lambda_\mu | u_\mu \rangle \otimes | b_\mu \rangle$. The states $| b_\mu \rangle = \sum_e V_{\mu e} | e \rangle$ are called bath states and, although they lie in a $D_E$-dimensional subspace, they span a $d_F$-dimensional subspace. The ground state of $\hat{H}$ is also the ground state of the subsystem Hamiltonian $\hat{H}_{emb} = \hat{P} \hat{H} \hat{P}$, where $\hat{P} = \sum_{\mu\nu} | u_\mu \rangle \langle u_\mu | \otimes | b_\nu \rangle \langle b_\nu |$ is the projection on a $2d_F$-dimensional subspace, drastically smaller than the original $D$-dimensional space. While this construction is formally exact and shows that the electronic structure of the entire system can be described exactly by that of a fragment and its surrounding bath, it is impractical because it assumes knowledge of the exact ground-state wavefunction and its Schmidt decomposition.

In practical implementations, DMET employs a mean-field approximation to the ground-state wavefunction~\cite{wouters2016practical}. For a closed-shell system of $N_{elec}$ electrons in $N_{orb}$ spatial orbitals, this is a Slater determinant of the form $| \Phi \rangle = \prod_{i\sigma} \crt{c_i\sigma} | \varnothing \rangle$, where $| \varnothing \rangle$ is the vacuum state and $\crt{i\sigma}$ creates a spin-$\sigma$ electron in the spatial orbital $|c_i \rangle = \sum_{p} C_{pi} | p \rangle$, with $i=1\dots N_{elec}/2$ and $|p \rangle$ a finite basis set of $N_{orb}$ orthonormal and spatially localized single-electron orbitals. Basis set elements are divided in fragment ($p \in F$) and environment ($e \notin F$) orbitals. Diagonalizing the environment-environment block of the one-particle reduced density matrix, $\rho_{e e^\prime} = \langle \Phi | \excitation{e}{e^\prime} | \Phi \rangle$ with $\excitation{e}{e^\prime} = \sum_\sigma \crt{e \sigma} \dst{e^\prime \sigma}$, yields a set of eigenpairs, $\sum_{e^\prime} \rho_{e e^\prime} W_{e^\prime \alpha} = W_{e\alpha} \delta_\alpha$.
The eigenvalues $\delta_\alpha=0$, $0 < \delta_\alpha < 2$, and $\delta_\alpha=2$ respectively correspond to virtual environment orbitals, so-called ``bath orbitals'', and occupied environment orbitals. The state $| \Phi \rangle$ can be factored in two terms, respectively containing occupied orbitals without overlap on $F$ (occupied environment) and with overlap on $F$ (linear combinations of fragment and bath orbitals). Since bath orbitals and local fragment orbitals are in general partially occupied in the DMET high-level wavefunction, the subsystem Hamiltonian has the form of a quantum chemistry active-space Hamiltonian~\cite{knizia2013density, wouters2016practical} where occupied and virtual environment orbitals are inactive orbitals. Denoting the electronic Hamiltonian of the full system as $\hat{H} = \sum_{pr} t_{pr} \, \excitation{p}{r} + \sum_{prqs} \frac{(pr|qs)}{2} \excitation{pq}{rs}$ with $\excitation{pq}{rs} = \sum_{\sigma\tau} \crt{p\sigma} \crt{q\tau} \dst{s\tau} \dst{r\sigma}$, the subsystem Hamiltonian takes the form
\begin{equation}
\label{eq:subsystem_hamiltonian}
\begin{split}
\hat{H}_{emb} &= \sum_{tu} \tilde{t}_{tu} \, \excitation{t}{u}
 + \sum_{prqs} \frac{(tu|vw)}{2} \, \excitation{tv}{uw} \;, \\
\tilde{t}_{tu} &= t_{tu} + \sum_{ee^\prime} \Big[ (tu|ee^\prime) - (te^\prime|eu) \Big] \tilde{\rho}_{ee^\prime} \;, \\
\end{split}
\end{equation}
where $tu$ range over fragment and bath orbitals (defining the subsystem, also called ``impurity'' in the DMET literature), and $\tilde{\rho}_{ee^\prime} = \sum_{\alpha\, :\,  \delta_\alpha=2} W_{e\alpha} W^*_{e^\prime \alpha}$ is the density matrix of the $N^{env}_{occ}$ occupied environment orbitals. The lowest-energy eigenvector (with $N_{elec} - 2 N^{env}_{occ}$ electrons) of the subsystem Hamilonian Eq.~\eqref{eq:subsystem_hamiltonian} can be computed with a high-level method like full configuration interaction (FCI)~\cite{knizia2012density, knizia2013density}, density matrix renormalization theory (DMRG)~\cite{zheng2016ground}, complete active space self-consistent field (CASSCF)~\cite{chen2014intermediate, pham2018can}, or CCSD~\cite{bulik2014electron}.

Typical DMET calculations employ multiple fragments, in which case multiple subsystem Hamiltonians are produced and diagonalized, yielding multiple one- and two-body reduced density matrices that are used to evaluate properties of the full system, e.g. the total energy and electron number~\cite{knizia2013density, wouters2016practical}. The partitioning into multiple overlapping fragments can lead to non-variationality of the DMET energy, and discrepancies between the target ($N_{elec})$ and DMET electron number.
To remedy this deficiency, we follow the strategy of ``one-shot'' DMET calculations~\cite{wouters2016practical}, where the subsystem Hamiltonians Eq.~\eqref{eq:subsystem_hamiltonian} are modified adding a global chemical potential (i.e. independent of fragment and orbital indices) for the local fragment orbitals, $\hat{H}_{emb} \to \hat{H}_{emb} - \mu_{\mathrm{glob}} \sum_{f \in F} \crt{f} \dst{f}$, optimized to ensure that DMET predicts the correct particle number. Optimization of chemical potential is done iteratively until the threshold is reached.

\subsection*{Sample-based quantum diagonalization}

SQD~\cite{kanno2023quantum,nakagawa2023adapt,robledo2024chemistry} is a quantum subspace method~\cite{motta2024subspace} that uses a quantum circuit $| \Phi_{\mathrm{qc}} \rangle$ to sample a set of computational basis states $\chi = \{ \bts{x}_1 \dots \bts{x}_d \}$ from the probability distribution $p(\bts{x}) = | \langle \bts{x} | \Phi_{\mathrm{qc}} \rangle |^2$ and a classical computer to solve the Schr\"{o}dinger equation in the subspace spanned by such computational basis states.
Since in the standard Jordan-Wigner (JW) mapping of fermions to qubits~\cite{ortiz2002simulating,somma2002simulating,somma2005quantum} a computational basis state parametrizes a Slater determinant, 
the Slater-Condon rules allow to efficiently compute the matrix elements $\langle \bts{x}_i | \hat{H} | \bts{x}_j \rangle$ within Davidson diagonalization.

In this work, we use SQD to perform conventional unfragmented active-space calculations~\cite{robledo2024chemistry, kaliakin2024accurate, barison2024ext-sqd} and to approximate the ground state of the DMET subspace Hamiltonians~\eqref{eq:subsystem_hamiltonian}. 
We sample computational basis states from the following truncated version of the local unitary cluster Jastrow (LUCJ) ansatz~\cite{motta2023bridging}
\begin{equation}
\label{eq:lucj}
| \Phi_{\mathrm{qc}} \rangle = e^{-\hat{K}_2} e^{\hat{K}_1} e^{i\hat{J}_{1}} e^{-\hat{K}_{1}} | {\bf{x}}_{\mathrm{RHF}} \rangle \;,
\end{equation}
where ${\hat{K}}_1$ and ${\hat{K}}_2$ are one-body operators, ${\hat{J}}_1$ is density-density operator, and $| {\bf{x}}_{\mathrm{RHF}} \rangle$ is the restricted closed-shell Hartree-Fock (RHF) state.
The parameters defining the LUCJ wavefunction~\eqref{eq:lucj} are derived from a classical restricted closed-shell CCSD, as detailed in Ref.~\citenum{robledo2024chemistry}.

On a noisy quantum device, computational basis states are sampled from a probability distribution $\tilde{p}(\bts{x})$ that unavoidably differs from $p(\bts{x})$ due to quantum noise. 
As a result, (i) particle number and total spin-$z$ may not be conserved, i.e. the sampled computational basis states may correspond to Slater determinants with incorrect number of spin-up and/or spin-down electrons, and 
(ii) the subspace spanned by the sampled computational basis states may not allow to produce eigenstates of the total spin operator $\hat{S}^2$. 
We emphasize that situations (i) and (ii) may occur also on noiseless quantum devices, respectively because the quantum circuit $\Psi$ may break particle-number and spin-$z$ symmetries (although this is not the case for LUCJ) 
and Slater determinants are generally not eigenstates of total spin.
To restore the broken particle-number and spin-$z$ symmetries, Ref.~\citenum{robledo2024chemistry} proposed an iterative self-consistent configuration recovery (S-CORE) procedure.
Each iteration of S-CORE has two inputs: a fixed set of computational basis states $\tilde{\chi}$ sampled from a quantum computer, and an approximation to the ground-state occupation number distribution $n_{p\sigma} = \langle \Psi | \crt{p\sigma} \dst{p\sigma} | \Psi \rangle$. In each iteration, we randomly flip the entries of the computational basis states in $\tilde{\chi}$ based on $n_{p\sigma}$ until particle number and spin-$z$ assume target values, thereby producing a new set $\tilde{\chi}_R$.
We then sample $K$ subsets (batches) from $\tilde{\chi}_R$, that we label $\tilde{\chi}_b$ with $b = 1 \dots K$. Each batch yields a subspace $S^{(b)}$ of dimension $d$~\cite{robledo2024chemistry}, in which we project the many-electron Hamiltonian as
\begin{equation}
\label{eq6}
\hat{H}_{S^{(b)}}=\hat{P}_{S^{(b)}}\hat{H}\hat{P}_{S^{(b)}}
\;,
\end{equation}
where the projector $\hat{P}_{S^{(b)}}$ is
\begin{equation}
\label{eq7}
\hat{P}_{S^{(b)}} =\sum_{ {\bf{x}} \in S^{(b)}} | {\bf{x}} \rangle \langle {\bf{x}} |
\;.
\end{equation}

We compute the ground-state wavefunctions and energies of Eq.~\eqref{eq6}, respectively $|\psi^{(b)} \rangle$ and $E^{(b)}$, and use the lowest energy across the batches, $\min_b E^{(b)}$, as the best approximation to the ground-state energy.
We use the wavefunctions $|\psi^{(b)} \rangle$ to update the occupation number distribution,
\begin{equation}
\label{eq8}
n_{p\sigma}=\frac{1}{K} \sum_{1\leq b\leq K } \langle \psi^{(b)} | \crt{p\sigma} \dst{p\sigma} | \psi^{(b)} \rangle
\;,
\end{equation}
and use it as an input in the next S-CORE iteration. At the first S-CORE iteration, we initialize $n_{p\sigma}$ by post-selecting~\cite{huggins2021efficient} measurement outcomes in $\tilde{\chi}$ with the correct particle number.

While S-CORE restores particle number and spin-$z$ conservation (and can be immediately generalized to any other symmetry operator having Slater determinants as eigenstates, e.g., molecular point-group symmetries in a basis of symmetry-adapted MOs),
it does not ensure conservation of total spin $\hat{S}^2$. To mitigate this limitation, in the construction of the subspaces $S^{(b)}$, we extend the set of configurations $\tilde{\chi}_b$ to ensure its closure under spin inversion symmetry as detailed in Ref.~\citenum{robledo2024chemistry}. For this reason, $d$ can be larger than $|\tilde{\chi}_b|$.

\subsection*{Computational details}

\begin{figure*}[t!]
\centering
\includegraphics[width=1.0\textwidth]{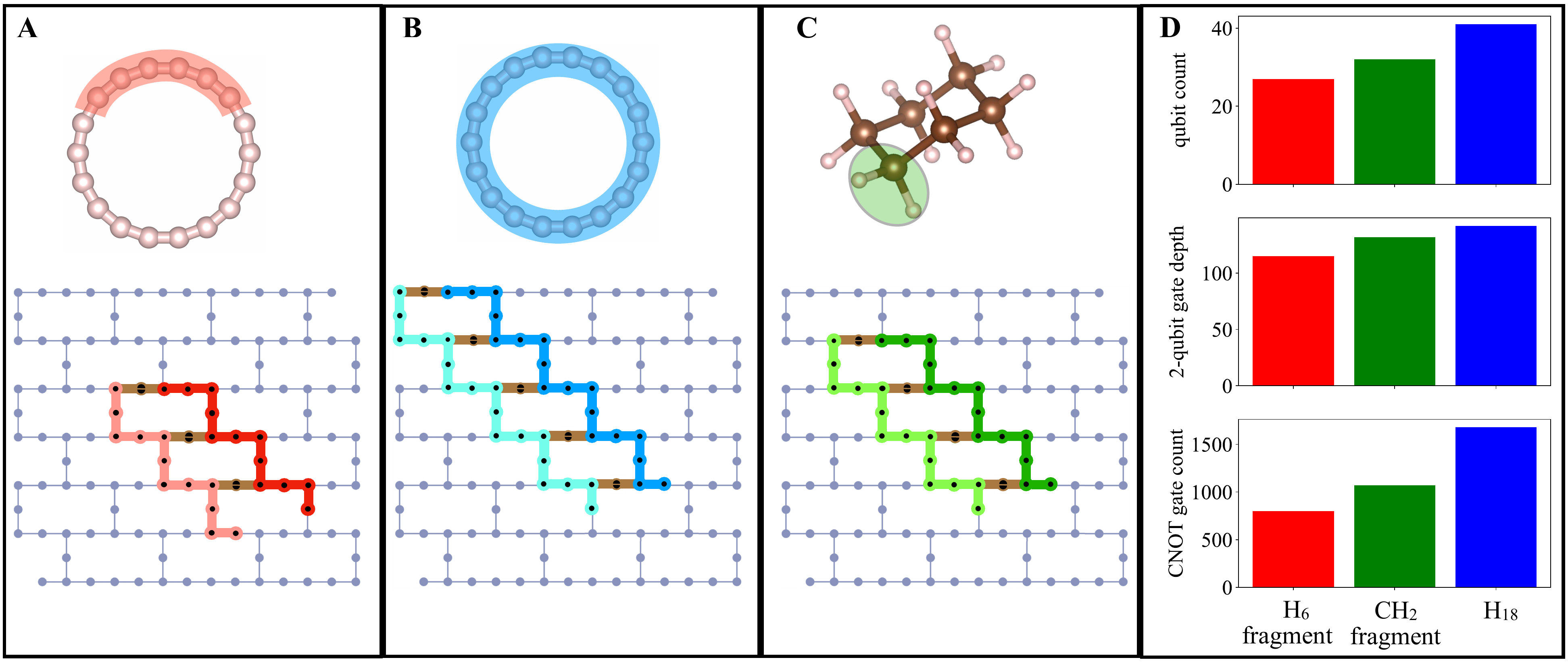}
\caption{\captiontitle{Overview of quantum circuits} (A-C) Qubit layouts of LUCJ circuits for: (A) DMET-SQD simulations of H$_{18}$ with H$_{6}$ fragments and (12e,12o) subsystems using 27 qubits of \device{cleveland}, (B) unfragmented H$_{18}$ simulations using 41 qubits, and (C) DMET-SQD simulations of cyclohexane with CH$_{2}$ fragments and (14e,14o) subsystems using 32 qubits. Colored highlights over the molecular structures indicate DMET fragments (red, blue, green for systems A, B, C respectively). The layout of \device{cleveland} is shown in gray, qubits used to encode occupation numbers of spin-up/down electrons are marked in light/dark colors, and ancilla qubits in brown. 
(D) Number of qubits, 2-qubit gate depth, and CNOT gate count of the LUCJ circuits of systems A, B, C (red, blue, green columns respectively).
}
\label{fig:layout}
\end{figure*}

\paragraph*{\textbf{Target applications}} We define the geometries of H$_{18}$ as $\bts{R}_k = (\rho \cos(k \Delta\theta) , \rho \sin(k \Delta\theta) , 0)$ with $k=1 \dots 18$ and $\Delta \theta = 2\pi/18$, i.e. along a circle of radius $\rho$, chosen so that the distance between adjacent atoms is $R =$ 0.70, 0.85, 1.00, 1.10, and 1.30 $\text{\AA}$ (see Fig.~\ref{fig:layout}). We compute the geometries of the chair, half-chair, twist-boat, and boat conformers of cyclohexane by performing a geometry optimization at the MP2/aug-cc-pVDZ level of theory using the ORCA software package~\cite{neese2022software}, and confirming the local minimum and transition-state character of each geometry by a vibrational frequency analysis.

\paragraph*{\textbf{DMET}} We perform DMET calculations as implemented in Tangelo 0.4.3~\cite{tangelo}, using PySCF 2.6.2~\cite{sun2020recent,sun2018pyscf,sun2015libcint}. In our DMET calculations, carried out at the STO-3G level of theory, we localize orbitals~\cite{pham2018can, wouters2016practical} with the meta-Lo\"{w}din scheme~\cite{fulde2017dealing}, and optimize the chemical potential $\mu_{\mathrm{glob}}$ with a convergence threshold of $\scientific{1}{-5}$.

\paragraph*{\textbf{LUCJ}} We generate the LUCJ circuits in Eq.~\eqref{eq:lucj} using the ffsim library~\cite{ffsim2024} interfaced with Qiskit 1.1.1~\cite{aleksandrowicz2019qiskit,javadi2024quantum}. 
We execute them on the \device{cleveland} quantum computer, using the qubit layouts shown in Fig.~\ref{fig:layout}A, \ref{fig:layout}B, and \ref{fig:layout}C for DMET calculations on H$_{18}$ with an H$_{6}$ fragment, unfragmented SQD calculations on H$_{18}$, and DMET calculations on cyclohexane with a CH$_2$ fragment, respectively. To mitigate quantum errors, we use gate twirling (but not measurement twirling) over random 2-qubit Clifford gates~\cite{wallman2016noise} and dynamical decoupling~\cite{viola1998dynamical,kofman2001universal,biercuk2009optimized,niu2022effects} as available through the SamplerV2 primitive of Qiskit's runtime library. The number of qubits, 2-qubit gate depth, and number of CNOT gates are listed, for the circuits executed in this work, in Fig.~\ref{fig:layout}D. Since we elected to perform ``one-shot'' DMET calculations, and to optimize $\mu_{\mathrm{glob}}$ using the configurations sampled at $\mu_{\mathrm{glob}}=0$ and simply repeating the S-CORE procedure for several values of $\mu_{\mathrm{glob}}$, the total number of circuits executed in this work is 3, 1, and 6 for the systems in Fig.~\ref{fig:layout}A, \ref{fig:layout}B, and \ref{fig:layout}C respectively.

\begin{table*}[t!]
\begin{tabular}{llllllll}
\hline\hline
system & fragment & AS & method & $|\tilde{\chi}_b|$ $[10^3]$ & $d$ $[10^5]$   &$d^\prime$ $[10^5]$     & $D_{\mathrm{AS}}$ $[10^5]$   \\
\hline
H$_{18}$       & H$_{6}$      & (12e,12o)     & DMET-SQD     & 1        &  5.0       &1.190           &8.5  \\
H$_{18}$       & H$_{6}$      & (12e,12o)     & DMET-SQD     & 2        &  6.9       &1.723           &8.5  \\
H$_{18}$       & H$_{6}$      & (12e,12o)     & DMET-SQD     & 3        &  8.0       &2.050           &8.5  \\
H$_{18}$       & H$_{6}$      & (12e,12o)     & DMET-SQD     & 4        &  8.3       &2.174           &8.5  \\
H$_{18}$       & H$_{6}$      & (12e,12o)     & DMET-SQD     & 5        &  8.5       &2.224           &8.5  \\
H$_{18}$       & unfrag. & (18e,18o)     & SQD          & 7        &  1163      &0.731           &23639 \\
H$_{18}$       & unfrag. & (18e,18o)     & SQD          & 8        &  1408      &0.825           &23639 \\
H$_{18}$       & unfrag. & (18e,18o)     & SQD          & 9        &  1712      &0.876           &23639 \\
H$_{18}$       & unfrag. & (18e,18o)     & SQD          & 10       &  2034      &0.980           &23639 \\
H$_{18}$       & unfrag. & (18e,18o)     & SQD          & 11       &  2346      &1.048           &23639 \\
H$_{18}$       & unfrag. & (18e,18o)     & SQD          & 12       &  2656      &1.075           &23639 \\
C$_{6}$H$_{12}$  & CH$_{2}$     & (14e,14o)     & DMET-SQD     & 6        &  86.7      &0.858           &118 \\
C$_{6}$H$_{12}$  & CH$_{2}$     & (14e,14o)     & DMET-SQD     & 7        &  88.7      &0.886          &118 \\
C$_{6}$H$_{12}$  & CH$_{2}$     & (14e,14o)     & DMET-SQD     & 8        &  97.2      &0.889           &118 \\
C$_{6}$H$_{12}$  & CH$_{2}$     & (14e,14o)     & DMET-SQD     & 9        &  99.9      &0.906           &118 \\
C$_{6}$H$_{12}$  & CH$_{2}$     & (14e,14o)     & DMET-SQD     & 10       &  103.8     &0.907           &118 \\
\hline\hline                       
\end{tabular}
\caption{\captiontitle{Details of SQD and DMET-SQD calculations} Fragments are defined by MOs localized spatially in the regions highlighted in Fig.~\ref{fig:layout}. We use the following abbreviations: ``unfrag.'' for unfragmented (i.e. conventional SQD), ``AS'' for active space (a subsystem, i.e. fragment+bath, in DMET-SQD calculations) and $D_{\mathrm{AS}}$ for Hilbert-space dimension as determined from the active-space orbitals and electrons in column 3. $\tilde{\chi}_b$ and $d$ is defined as in the main text, and $d^\prime$ is the number of configurations in an SQD wavefunction $| \psi^{(b)} \rangle = \sum_{m=1}^d c_m | \bts{x}_m \rangle$ with amplitudes $|c_m|^2 \geq \scientific{1}{-8}$. The values of $d$ and $d^\prime$ are computed for H$_{18}$ at bondlength 1.0 $\text{\AA}$ and for the chair conformation of cyclohexane, and more extensive studies are reported in the Appendix.
}
\label{table:sqd}
\end{table*}

\paragraph*{\textbf{SQD}} We perform SQD calculations using the implementation in the GitHub repository Ref.~\citenum{sqd_addon}. 
In DMET-SQD simulations, we perform 10 and 3 iterations of S-CORE, respectively prior and during the optimization of the chemical potential. In unfragmented SQD calculations, we perform 10 iterations of S-CORE. The details of SQD and DMET-SQD calculations performed in this work are listed in Table~\ref{table:sqd}.

\paragraph*{\textbf{Classical benchmarks}} To assess the accuracy of DMET-SQD calculations, we perform unfragmented calculations using: CCSD, CCSD with perturbative triples (CCSD(T))~\cite{bartlett2024perspective} as implemented in PySCF 2.6.2, and heat-bath configuration interaction (HCI)~\cite{holmes2016heat,holmes2016efficient,smith2017cheap,sharma2017semistochastic} as implemented in the SHCI-SCF 0.1 interface between PySCF 2.6.2 and DICE 1.0. In HCI simulations, we fix the parameter $\varepsilon_1$ (controlling which determinants are included in the variational wavefunction~\cite{holmes2016heat}) to $\scientific{5}{-6}$ during initial variational steps and $\scientific{1}{-6}$ during later variational steps.

\vspace{2em}
\tocless\section{Results}

We now present results for a ring of 18 hydrogen atoms, closely related to the hydrogen chain described in Ref.~\citenum{hachmann2006multireference} and studied at finite lengths and finite basis sets by several groups~\cite{al2007bond,tsuchimochi2009strong,sinitskiy2010strong,lin2011dynamical,stella2011strong,mazziotti2011large}
both as a benchmark for numerical methods in presence of electronic correlation of varying nature and strength~\cite{motta2017towards} and a model system to investigate a variety of physical phenomena~\cite{motta2020ground}.

\begin{figure}[!ht]
     \centering
     \includegraphics[width=\columnwidth]{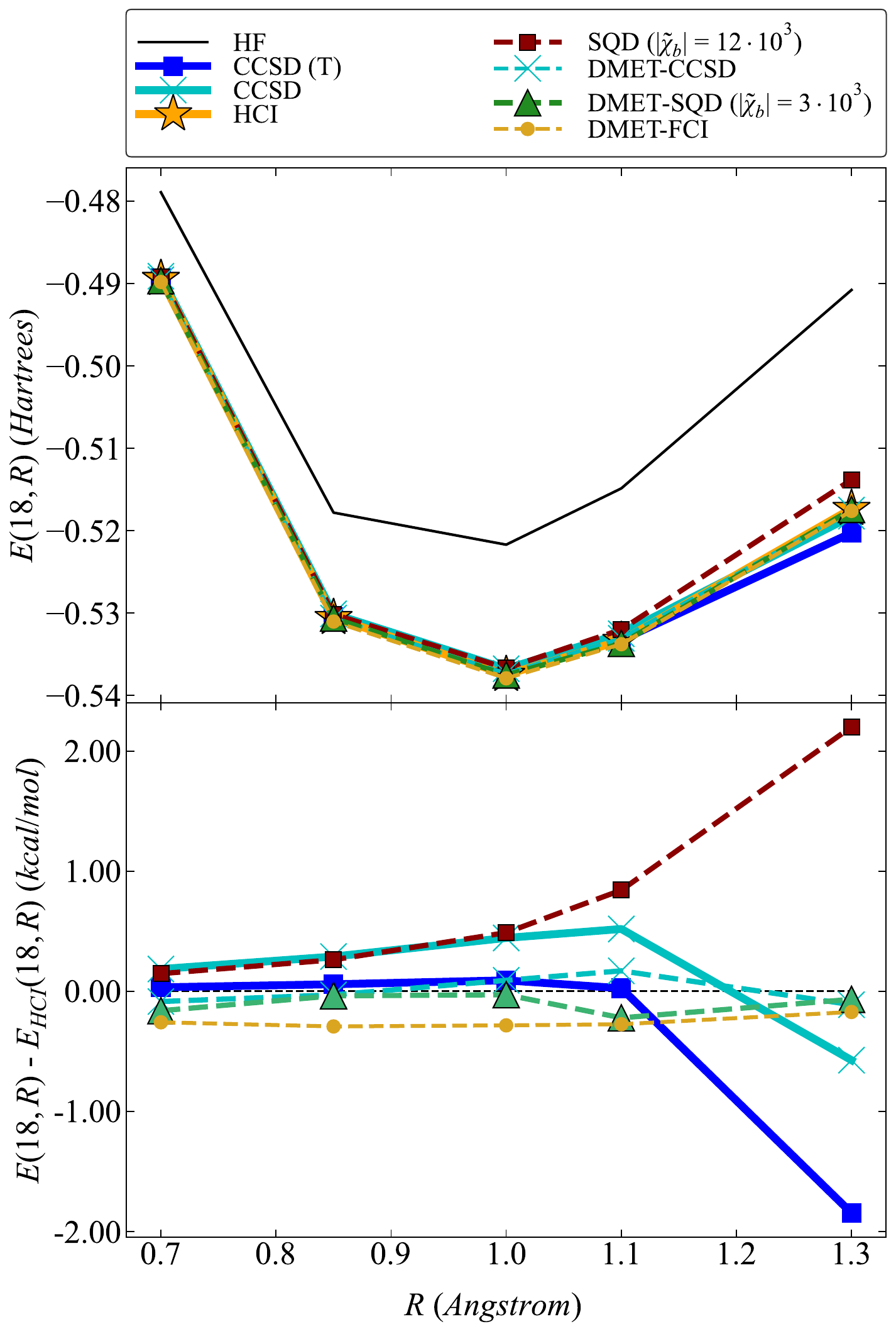}
     \caption{\captiontitle{Potential energy curve of a hydrogen ring} Top: ground-state energy per atom along the symmetric expansion of a ring of 18 hydrogen atoms using various unfragmented (RHF, CCSD, CCSD(T), SQD, HCI) methods and DMET with CCSD, SQD, and FCI as subsystem solver. Bottom: deviation between HCI and other potential energy curves. Energies per atom are shown.}
     \label{fig: Figure 2}
\end{figure}

In Fig.~\ref{fig: Figure 2}, we show the potential energy curve of the ring from unfragmented restricted closed-shell HF, CCSD, CCSD(T), SQD, and HCI methods (top) and DMET with CCSD, SQD, and FCI as subsystem solver (bottom). 
Correlated methods agree qualitatively for $R \leq 1.0 \,\text{\AA}$ and their potential energy curves visibly depart from each other for $R \geq 1.1 \text{\AA}$, as the simultaneous breaking of multiple bonds leads to stronger electronic correlation effects.
The different behavior of correlated methods is analyzed in more detail in the bottom panel of Fig.~\ref{fig: Figure 2}, where we show deviations between computed energies and HCI, taken as reference.
As $R$ increases restricted CCSD and CCSD(T) underestimate the ground-state energy by $\sim$2 kcal/mol per atom, a phenomenon well-known~\cite{bulik2015can, motta2017towards} to occur in the presence of static electronic correlation.
On the other hand SQD overestimates it by $\sim$2.5 kcal/mol per atom: as $R$ increases, the ground-state wavefunction acquires multireference character, requiring more configurations for accurate energy estimates, and highlighting the sensitivity of SQD to several algorithmic elements including the LUCJ circuit, its parametrization via CCSD amplitudes, and device noise (affecting e.g. the quality of the occupation number distribution used in the first iteration of S-CORE). 

DMET-CCSD and DMET-SQD yield potential energy curves in better agreement with HCI than equivalent unfragmented calculations and with considerably lower non-parallelity error (beneficial in the calculation of energy differences and derivatives) and sub-kcal/mol non-variationality biases.
The observed improvements are accounted for by the smaller size of DMET subsystems as compared to the entire hydrogen ring, which alleviates the breakdown of CCSD in DMET-CCSD calculations, and the impact of inefficient configuration sampling in DMET-SQD calculations (the ratio between the number $d^\prime$ of configurations with coefficients above a given threshold, and the total number $d$ of configurations in the SQD wavefunction is higher for DMET-SQD than for unfragmented SQD, see Table~\ref{table:sqd} and the Appendix).

\begin{figure}[!ht]
     \centering
     \includegraphics[width=\columnwidth]{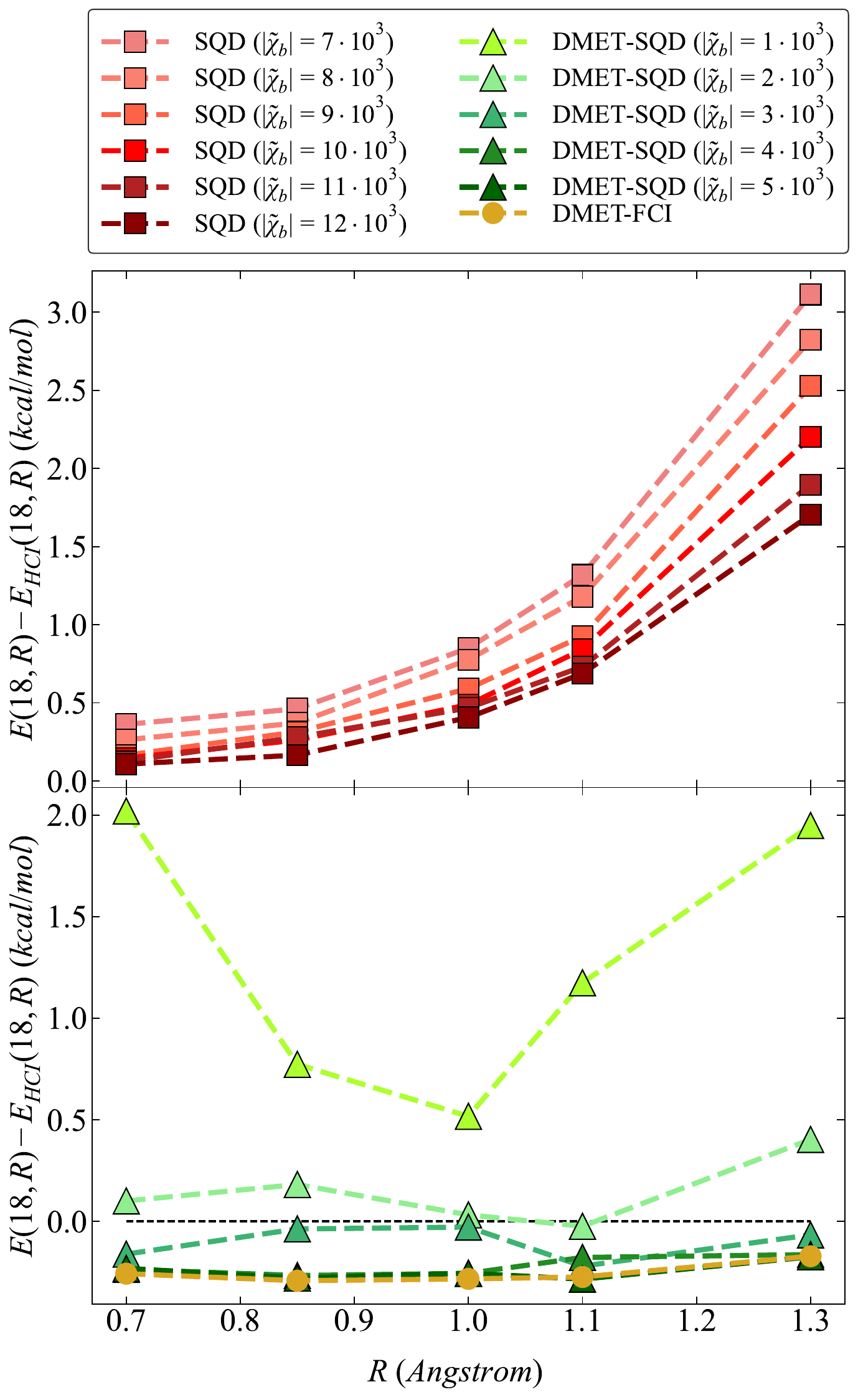}
     \caption{\captiontitle{SQD performance versus number of configurations} Deviations from HCI for unfragmented SQD (top) and DMET-SQD energies (bottom), along the symmetric expansion of a ring of 18 hydrogen atoms, for several values of $|\tilde{\chi}_b|$.}
     \label{fig: Figure 3}
\end{figure}

In Fig.~\ref{fig: Figure 3}, we explore the behavior of SQD and DMET-SQD calculations as a function of the number of batch configurations $|\tilde{\chi}_b|$. As naturally expected, increasing $|\tilde{\chi}_b|$ leads to increasing $d$ (see also Table~\ref{table:sqd}) and thus to a closer agreement with HCI and DMET-FCI, respectively.
Increasing $|\tilde{\chi}_b|$ also reduces fluctuations in SQD energies caused by the randomness of sampled configurations, which are particularly visible in the bottom panel of Fig.~\ref{fig: Figure 3} (light green curve, $|\tilde{\chi}_b|=\scientific{1}{3}$).
The rate at which SQD and DMET-SQD results approach HCI and DMET-FCI counterparts is determined by $|\tilde{\chi}_b|$ through the ratio between $d$ and the dimension of the active space and through the ratio between $d^\prime$ and $d$, which is lower for unfragmented H$_{18}$ calculations (see Table~\ref{table:sqd} and the appendix), suggesting inefficient sampling.

\begin{figure}[!ht]
     \centering
     \includegraphics[width=\columnwidth]{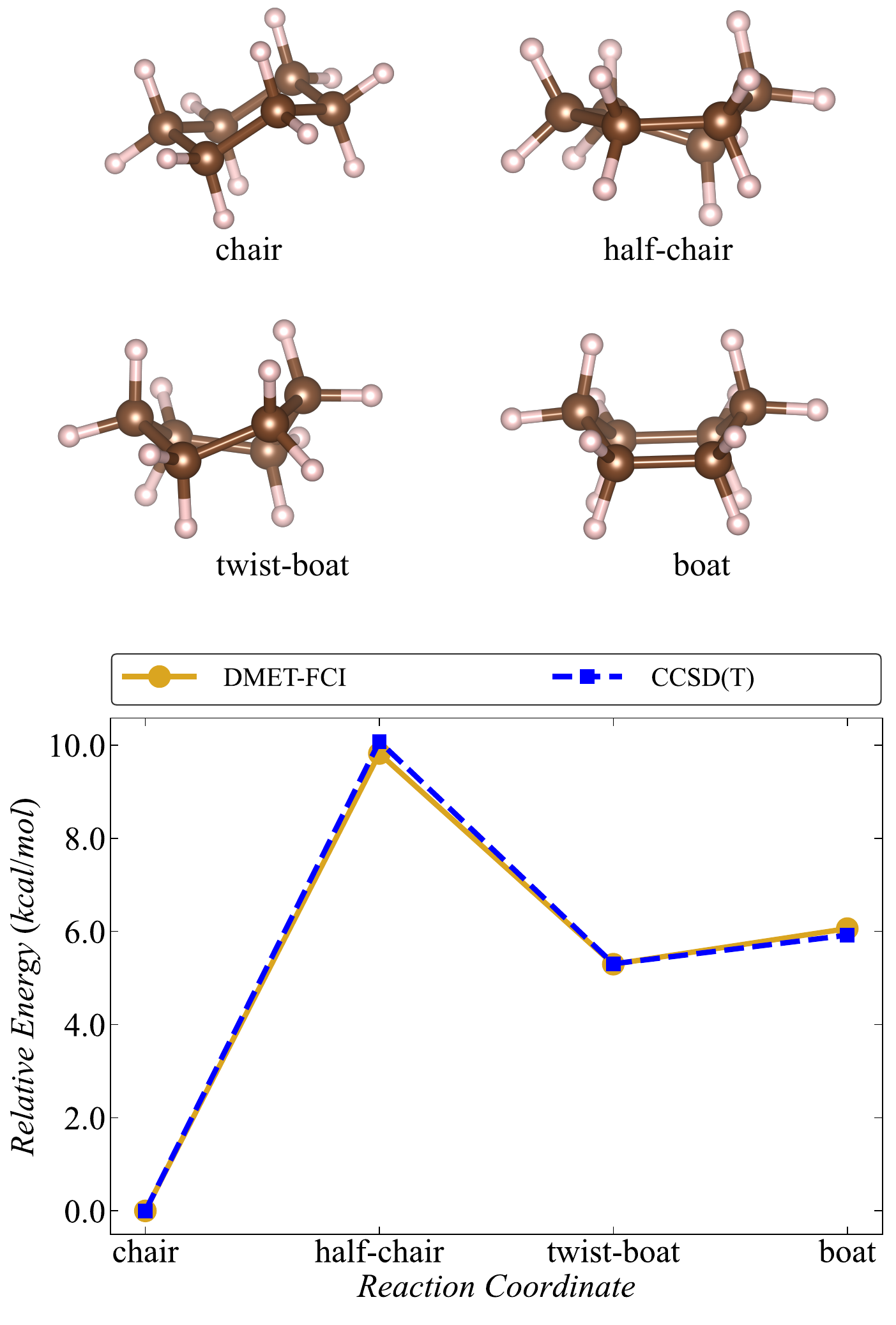}
     \caption{\captiontitle{Conformations of cyclohexane} Schematic representations of the chair, half-chair, twist-boat, and boat conformations of cyclohexane (top) and their energies relative to the chair using DMET-FCI and unfragmented CCSD(T) (bottom).}
     \label{fig: Figure 4}
\end{figure}

\begin{figure}[!ht]
     \centering
     \includegraphics[width=\columnwidth]{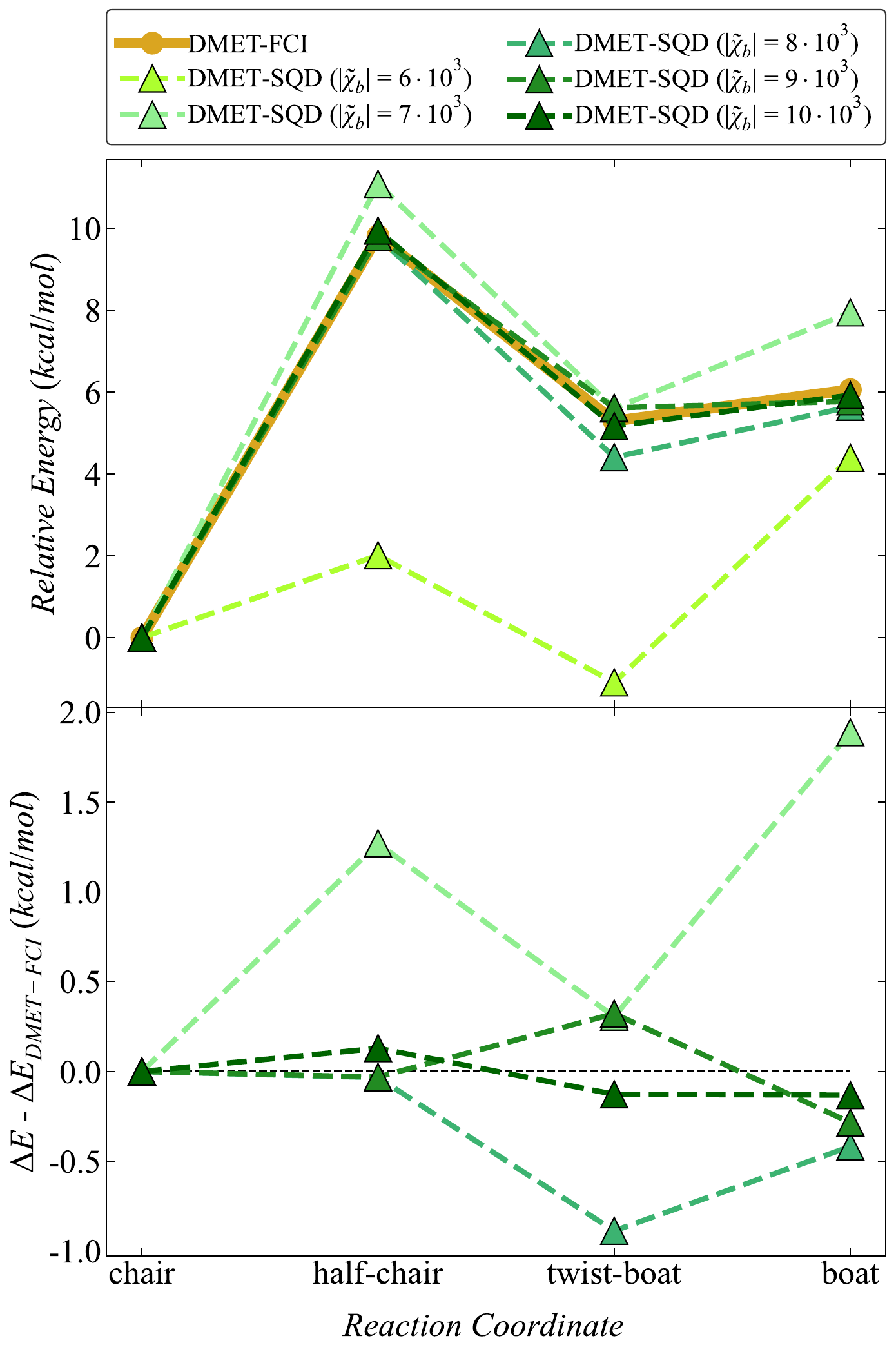}
     \caption{\captiontitle{DMET-SQD calculations on cyclohexane} Top: comparison between DMET-FCI and DMET-SQD for several values of $|\tilde{\chi}_b|$. Bottom: deviations of DMET-SQD from DMET-FCI.}
     \label{fig: Figure 5}
\end{figure}

We now turn our attention to cyclohexane. This compound can adopt several conformations, illustrated in Fig.~\ref{fig:  Figure 4}, owing to the flexibility of its carbon-carbon single bonds: 
(i) in the lowest-energy ``chair'' conformation, C atoms are positioned so that C-C ring bonds assume energetically favorable angles (thereby minimizing angle strain) and C-H ring bonds are staggered (eliminating torsional strain),
(ii) in the less stable ``boat'' conformation, two C atoms are at the same height, creating steric strain (and hindrance) due to the proximity of two H atoms bound to them (above the boat) and creating torsional strain because eight H atoms (below the boat) are forced into eclipsed positions,
(iii) in the ``twist-boat'' conformation, a slight twisting deformations from the ``boat'' reduces steric strain by moving H atoms ``above the boat'' far apart, and eclipsing interactions by moving H atoms ``below the boat'' into largely (but not completely) staggered positions,
(iv) in the ``half-chair'' conformation, cyclohexane assumes a partially planar shape at the cost of significant ring strain, as in the planar portion the C-C bond angles are forced to energetically unfavorable angles and the corresponding C-H bonds are fully eclipsed.

Computing energy differences between cyclohexane conformations in a qualitatively (i.e. correctly ranking) and quantitatively accurate way is an important methodological test, particularly because these energy differences are of the order of a few kcal/mol, and a necessary step towards richer and more realistic studies.
In Fig.~\ref{fig:  Figure 4}, we compare DMET-FCI with unfragmented CCSD(T) energies of cyclohexane conformations, relative to the chair. As seen, the fragmentation approximation underlying DMET affects energy differences rather modestly, well below 1 kcal/mol.
In Fig.~\ref{fig:  Figure 5} (top), we compare DMET-FCI and DMET-SQD energy differences for several values of $|\tilde{\chi}_b|$, showing deviations between DMET-FCI and DMET-SQD for each conformation in Fig.~\ref{fig:  Figure 4} (bottom).
For $|\tilde{\chi}_b| \geq \scientific{8}{3}$, deviations from DMET-FCI are mostly within 1 kcal/mol and conformations are ordered correctly by DMET-SQD. 
For $|\tilde{\chi}_b|=\scientific{6}{3}$, on the other hand, the ordering of conformations is incorrect, with twist-boat predicted to be the most stable geometry and half-chair predicted to be more stable than boat.

\vspace{2em}
\tocless\section{Conclusion}

In this study, we presented the first DMET calculations using the quantum computing SQD method as a subsystem solver. 
We tested the DMET+SQD combination by computing the the ground-state potential energy curve of a ring of 18 hydrogen atoms, a standard benchmark of first-principle electronic structure methods, and the relative energies of the conformers of cyclohexane, a use case of more practical significance to organic chemistry.
The relative energies of these conformations do not results from the breaking and formation of chemical bonds, but from the delicate balance of electrostatic interactions between molecular moieties in different geometries. In this aspect, the calculations performed in this work can be considered a severe test of DMET-SQD.

The use of SQD as a subsystem solver allowed the quantum computation of subsystems with more electrons and orbitals than previously possible, by executing fewer and larger quantum circuits (as documented in Fig.~\ref{fig:layout}) on a quantum processor assisted by classical HPC resources carrying out pre-, peri-, and post-processing operations (respectively preliminary mean-field calculations and subsystem Hamiltonian construction, S-CORE and the associate subspace diagonalization, and computation of system properties such as particle number and ground-state energy by the collation of results from individual subsystems).
In addition, the use of SQD as a subsystem solver allows for higher accuracy and precision than previously possible in studies involving present-day quantum computers, due to the noise-resilience properties of the algorithm~\cite{robledo2024chemistry}.

It is important to continue to expand the application of DMET-SQD to more complex instances of the electronic structure problem. Continued development in error rates of quantum computers, error mitigation techniques, and construction and optimization of quantum circuits for SQD applications have the potential to improve the efficiency of sampling computational basis states, leading to more compact eigenvalues problems. This would in turn improve the accuracy and time to solution of DMET-SQD, further extending the range of accessible applications and/or the reliance on computationally expensive HPC peri-processing (especially matrix diagonalization).
Further studies building on the work done here, involving chemical species of greater relevance to organic and biological chemistry, and non-minimal basis sets necessary for qualitatively correct and quantitatively accurate results, would also be very desirable.

More generally, our study serves as a proof-of-concept for the concerted use of quantum and classical computers, as complementary elements of QCSC algorithms and architectures, as an innovative and promising mode of attack of correlated many-electron systems by first-principle electronic structure methods, with the ultimate goal to enhance molecular design platforms.

\vspace{2em}
\tocless\acknowledgements
The authors gratefully acknowledge financial support from the National Science Foundation (NSF) through CSSI Frameworks Grant OAC-2209717 and from the National Institutes of Health (Grant Numbers GM130641). The authors also thank M. Rossmannek and I. Tavernelli for useful feedback on the manuscript.

\newpage
\tocless\section{Additional data}

\begin{figure}[t!]
     \centering
     \includegraphics[width=\columnwidth]{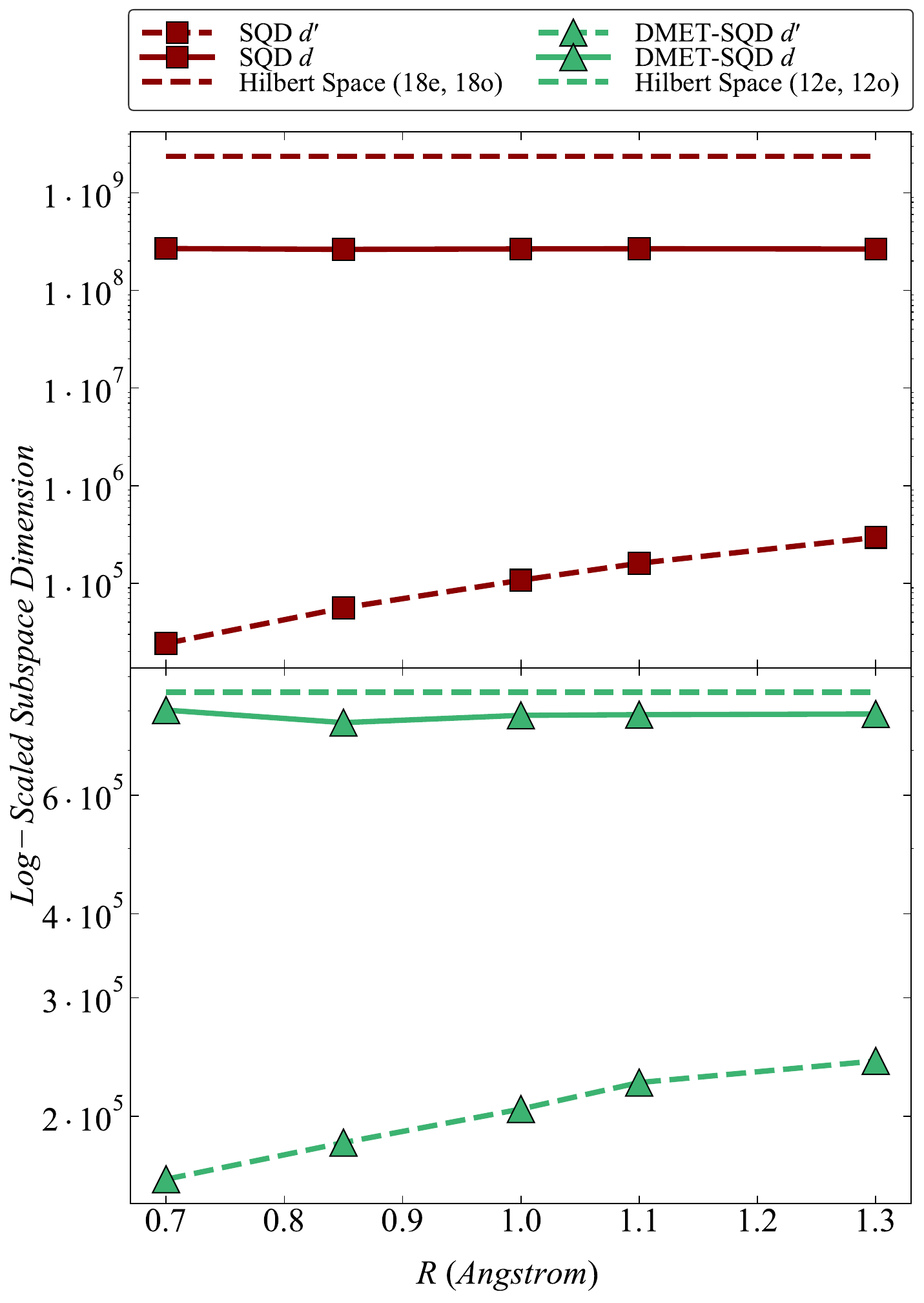}
     \caption{\textbf{Sparsity of SQD wavefunctions for a hydrogen ring} Hilbert space dimension (dashed line), number $d$ of configurations in the SQD wavefunction (symbols connected by a solid line) and number $d^\prime$ of such configurations with coefficients above $1\cdot 10^{-8}$ in absolute value squared (symbols connected by a dashed line), for a ring of 18 hydrogen atoms as a function of bondlength, studied with SQD (top, red squares) and DMET-SQD (bottom, green triangles).}
     \label{fig: Figure 6}
\end{figure}

\begin{figure}[ht!]
     \centering
     \includegraphics[width=\columnwidth]{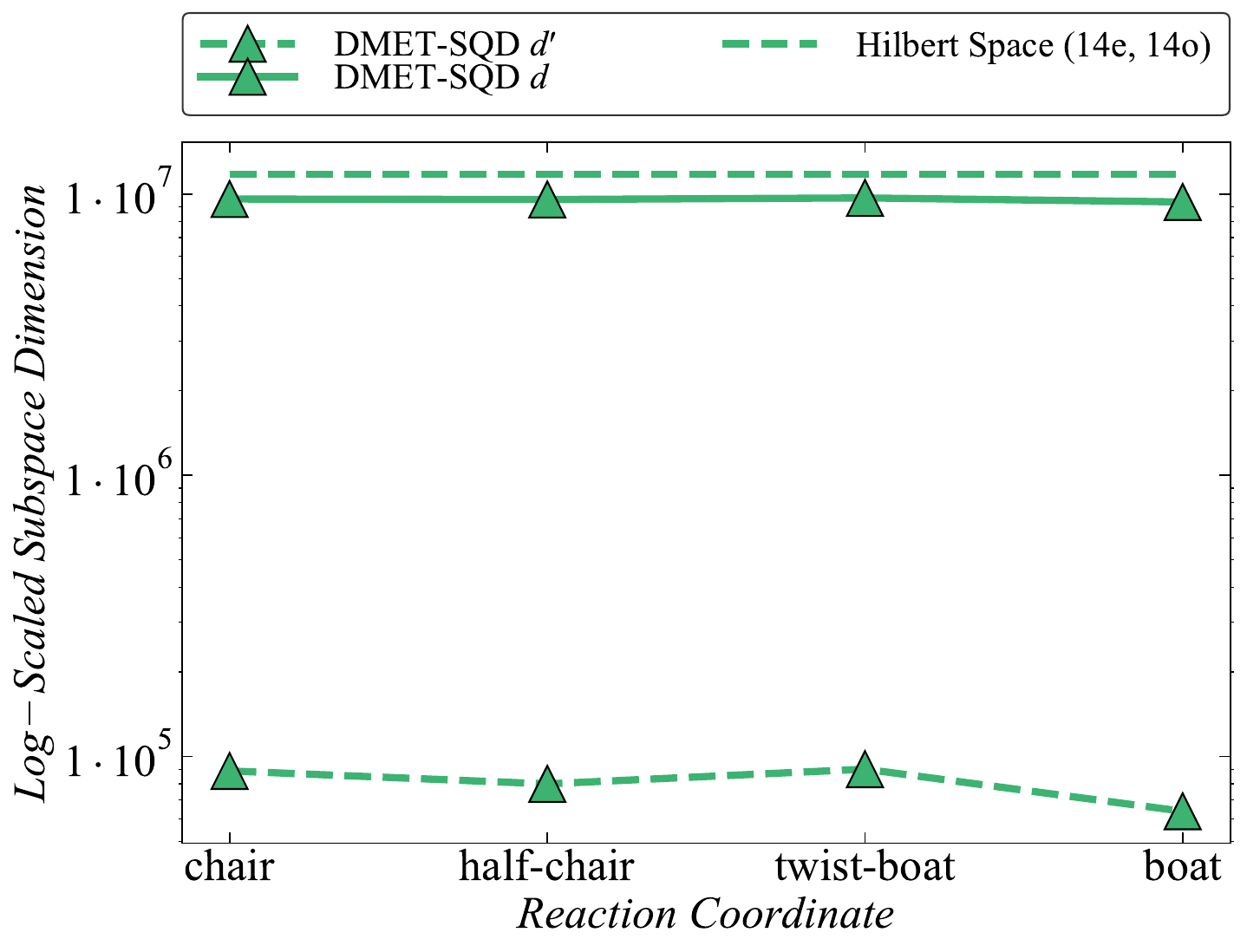}
     \caption{\textbf{Sparsity of SQD wavefunctions for cyclohexane} Hilbert space dimension (dashed line), number $d$ of configurations in the SQD wavefunction (triangles connected by a solid line) and number $d^\prime$ of such configurations with coefficients above $1.0 \cdot 10^{-8}$ in absolute value squared (triangles connected by a dashed line), for cyclohexane conformations studied with DMET-SQD.}
     \label{fig: Figure 7}
\end{figure}

In Figs.~\ref{fig: Figure 6} and \ref{fig: Figure 7}, we show the number of configurations in an SQD wavefunction $\left| \psi^{(b)} \rangle = \sum_{m=1}^d c_m \right| \bts{x}_m \rangle$, labeled $d$, the number of such configurations with $|c_m|^2 \geq 1 \cdot 10^{-5}$ labeled $d^\prime$, and the Hilbert space dimension, for H$_{18}$ as a function of bondlength using $\left| \tilde{\chi}_b \right| = 12\cdot 10^{3}$ and for cyclohexane as a function of reaction coordinate using $\left| \tilde{\chi}_b \right| = 8\cdot 10^{3}$ respectively.

In Fig.~\ref{fig: Figure 6}, note how $d$ is relatively close to the Hilbert space dimension and roughly independent of $R$, whereas $d^\prime$ increases with $R$, albeit at different rates for SQD and DMET-SQD.

\newpage
\bibliography{DMET_SQD}

\end{document}